\documentclass[a4paper]{spie}

\usepackage[sc, hang]{caption}
\usepackage{graphicx}
\usepackage{graphics}
\usepackage{makeidx}
\usepackage[pass]{geometry}
\usepackage{booktabs}
\usepackage{marvosym}
\usepackage[hang]{subfigure}
\usepackage{SIunits}
\usepackage{rotating}
\usepackage{csquotes}
\usepackage{url}
\usepackage{multicol}
\usepackage{multirow}
\usepackage{overpic}
\usepackage{amsmath}
\usepackage{commath}
\usepackage{bm}
\usepackage{rotating}
\usepackage{psfrag}
\usepackage{parallel}
\usepackage{fancyhdr}
\usepackage[breaklinks]{hyperref}

\graphicspath{{immagini/}{/Users/tmazzoni/Documents/latex/immagini/istituzioni/}}  
\usepackage[english]{babel} 
\usepackage{eso-pic,graphicx}
\usepackage{enumerate}
\usepackage{psfrag}
\usepackage{booktabs}
\usepackage{hyperref}      % - per collegamenti ipertestuali. Deve essere caricato per ultimo ma prima di ``cite''
\hypersetup{               % - specifica le caratteristiche di ``hyperref''
   pdftitle={ocamSPIE2018},    % title
   pdfauthor={Tommaso Mazzoni},  % author
   colorlinks=true,      % disables boxes surounding links
   linkcolor=black,      % normal internal links
   anchorcolor=black,    % anchor text
   citecolor=black,      % bibligraphical citations in text
   filecolor=black,      % URLs which open local files
   menucolor=black,      % Acrobat menu items
   pagecolor=black,      % links to other pages
   urlcolor=blue,         % URLs
   linktocpage=true
}

\graphicspath{{/Users/tmazzoni/Documents/latex/articoloSPIE2018/immagini/}
  {/Users/tmazzoni/Documents/latex/grafici/argos/camera/cosmic/}
  {/Users/tmazzoni/Documents/latex/immagini/istituzioni/}}

\title{EMCCD for Pyramid wavefront sensor: laboratory characterization} 

\author[a]{Guido Agapito}
\author[a]{Tommaso Mazzoni}
\author[a]{Fabio Rossi}
\author[a]{Alfio Puglisi}
\author[a]{Cedric Plantet}
\author[a]{Enrico Pinna}
\affil[a]{INAF Osservatorio Arcetri, L. Enrico Fermi 5, 50125 Firenze, Italy}
\authorinfo{Further author information:\\Guido Agapito, \Letter
 \hspace{0.5ex} agapito@arcetri.astro.it\\}

\begin{document} 

  \maketitle

  \begin{abstract}
    Electro-Multiplying CCDs offer a unique combination of speed, sub-electron noise and quantum efficiency. These features make them extremely attractive for astronomical adaptive optics. The SOUL project selected the Ocam2k from FLI as camera upgrade for the pyramid wavefront sensor of the LBT SCAO systems. Here we present results from the laboratory characterization of the 3 of the custom Ocam2k cameras for the SOUL project. The cameras showed very good noise ($0.4e^-$ and $0.4-0.7e^-$ for binned modes) and dark current values ($1.5e^-$). We measured the camera gain and identified the dependency on power cycle and frame rate. Finally, we estimated the impact of these gain variation in the SOUL adaptive optics system. The impact on the SOUL performance resulted to be negligible.
    
  \end{abstract}


\section{INTRODUCTION}
\label{sec:}

Wavefront sensors for astronomical adaptive optics require fast and low noise
cameras with hundreds by hundreds of pixels. Following these requirements, in the last 10 years, the top class telescopes started to use a new class of CCDs, 
the Electron-multiplying CCD. These new devices have been introduced in AO systems\cite{doi:10.1117/12.2233319,2014SPIE.9148E..5CE,AO4ELT3.13248} with Shack Hartmann\cite{hartmann} and Pyramid WaverFront Sensors\cite{pyramid} (PWFS).

Ocam2k by First Light Advanced Imagery
\cite{2011PASP..123..263F,doi:10.12839/AO4ELT3.15019}
(FLI) is an EMCCD
with sub-electron read out noise, implementing the CCD220 by e2v \cite{doi:10.1117/12.788009, 2006ASSL..336..321D}.

The aim of this paper is to present the results of the laboratory characterization
of 3 of the 5 Ocam2k cameras that will be implemented on the PWFS of the Single
conjugated adaptive Optics Upgrade for LBT (SOUL) \cite{SOULSPIE2016}. 
In fact, the SOUL project will upgrade the four existing SCAO systems of LBT
\cite{2010ApOpt..49..115H,2011SPIE.8149E...1E}, and Ocam2k
has been selected to replace the Little Joe CCD39 camera by SciMeasure
\cite{2000SPIE.4007..481D}
currently operating at LBT.

FLI customized these Ocam2k cameras to match SOUL specifications:
the mechanical layout has been expressly designed for the implementation on the FLAO WFS and 2 extra read-out modes added (on-chip binning mode 3x3 and 4x4).

Ocam2k camera has a negligible RON and larger format with respect to Little Joe CCD39:
this allows to improve contrast on bright magnitude guide star
thanks to an higher sampling of the pupil and to push the limiting magnitude
of about 2 magnitudes thanks to the lower measurement noise. Finally, the turbulence and telescope vibration will be rejected with higher efficiency thanks to the higher frame rate \cite{SOULSPIE2016}.
%\\
%The use of OCAM2k for the PWFS is currently considered even in the content of SCAO
%system for the next generation of 30-40m diameter telescopes.
%This paper will report some consideration about them,
%taking into account the presented lab results.

\section{Analysis of dark frames}
\label{sec:Dark}

In this section, we describe the results obtained analyzing frames without 
light (metallic screw cap on). These measurements are used to compute
Read-Out Noise, Dark Current and to evaluate Bias stability. 
Note that, we consider the value for the on-chip 
multiplication gain as set on the camera interface and the nominal one for
the CCD gain ($g = 1/30$), for all pixels.
We acquired 1 set of 5999 frames (the buffer size of the frame-grabber) 
for each combination of the following parameters:
\begin{itemize}
\item Acquisition mode = 1 (full frame), 2 (crop), 3 (binning 2x2), 4 (binning
3x3) and 5 (binning 4x4)
\item Frame rates = 100, 200, 300, 500, 1000, 1500, 2060Hz
\item On chip multiplication gain $m = 1$ and $m = 600$.
\end{itemize}
We repeated these acquisitions after few hours and a full power cycle of the camera for a total of 156 sets.

\subsection{Read-Out Noise}
\begin{figure}[h!]
\centering
\subfigure[Average dark frame at 2067 fps (6000 frames).\label{fig:aveDark}]
{\includegraphics[width=0.45\columnwidth]{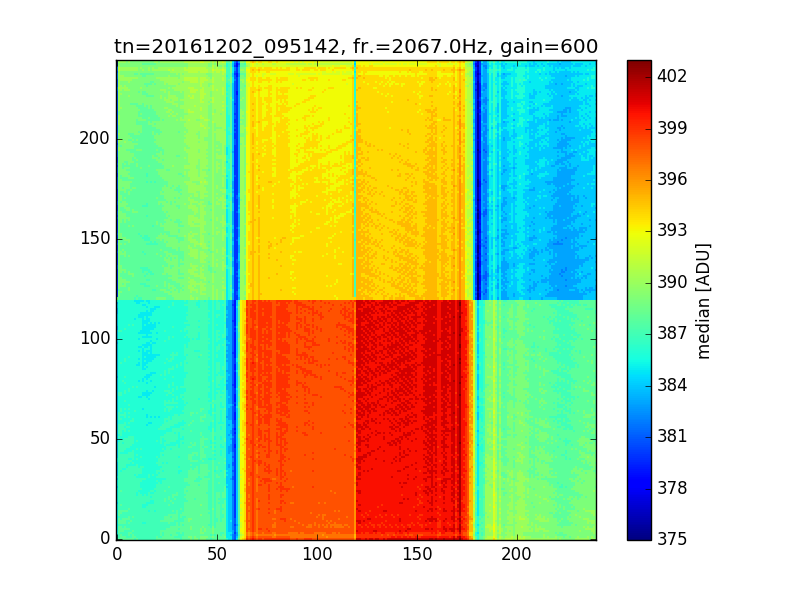}}
\subfigure[Counts Power Spectral Density for different frame rate speeds. Dark frames..\label{fig:psd}]
{\includegraphics[width=0.45\columnwidth]{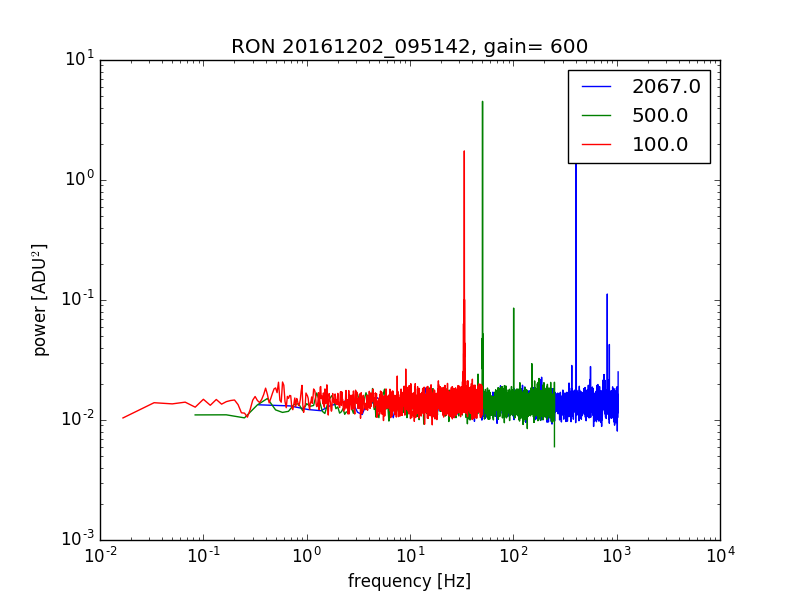}}
\caption{Output of dark frames analysis.}
 \label{fig:darkAnalysis}
\end{figure}

An example of dark frame is reported in Figure \ref{fig:aveDark}. The noise does not present 
the typical white spatial distribution due to the ADC read-out. High spatial 
frequency patterns are present over the entire frame. The averaged temporal 
spectrum presents clear structures too as can see in Figure \ref{fig:psd}. 

Read-Out Noise (RON), is estimated pixel by pixel from the temporal standard
deviation at the maximum frame rate, $f_{MAX}$:
\begin{equation} 
R_e= \sigma(f_{MAX})= \sqrt{\frac{\sum_{i=1}^N (p_i-\bar{p})^2}{N}}
\end{equation}
where $p_i$ is the pixel value expressed in $e^-=\frac{c_i}{mg}$,
$c_i$ are the pixel counts, $m$ is the on-chip multiplication
gain, $g$ is the CCD gain expressed in $ADU/e^-$,
$\bar{p}$ is the average pixel counts value,
$i$ is the frames index and $N=5999$.\\
We measured $R_e$ in with on-chip multiplication gain $m = 600$.
The results are summarized in Table \ref{tab:ron}. Notice that, as said before, in
this analysis, we considered the value of $m$ as set in the camera interface (600) and the nominal one for the
CCD gain (g = 0.03$ADU/e^-$), for all pixels.
The obtained values of noise are in good agreement with the ones provided 
by e2v and FLI, as can be seen in Table \ref{tab:ron}. As
expected by FLI, the on-chip binning slightly increases the RON. 

All the measured values are compliant with the specifications:
RON$<0.5 e^-$ with $m=600$ without binning and
RON$<1e^-$ with $m=600$ for binned modes.  

\begin{table}[h]
\centering
\caption{Read-Out noise for the three cameras. 
Third row reports the values given by e2v for m= 1000. Fourth and fifth rows are
the values from First Light for binning 1x1 and 2x2 respectively. The values in
the test in Arcetri labs are shown in the last 4 rows.  }\label{tab:ron}
\begin{tabular}{llccc}
	\toprule
	& \bf{Binning} & & \bf{$R_e$ [e$^-$]} & \\
	\bf{Camera no.} & & 
   	 \bf{39} &
   	 \bf{40} &
   	 \bf{41}\\
    \midrule
    \multicolumn{2}{l}{\bf{e2v (m=1000, 1274fps)}} & 0.295 & 0.332 & 0.347 \\
  	\multirow{2}{*}{\bf{First Light (m=600, 2000fps)}} & bin 1x1  
  	& 0.384 & 0.344 & 0.344  \\
  	  & bin 2x2  &  0.520 & 0.431 & 0.479   \\
  	\bf{OAA (m=600, 2066fps)} & bin 1x1  &  0.372 & 0.343 & 0.339   \\
  	\bf{OAA (m=600, 3620fps)} & bin 2x2  &  0.489 & 0.424 & 0.465   \\
  	\bf{OAA (m=600, 4890fps)} & bin 3x3  &  0.633 & 0.605 & 0.556   \\
  	\bf{OAA (m=600, 5900fps)} & bin 4x4  &  0.739 & 0.671 & 0.668   \\
    \bottomrule
  \end{tabular}
\end{table}

\subsection{Dark Current}

Dark Current, $D_e$, is estimated from the linear relationship between
integration time and temporal variance:
\begin{equation}
\sigma^2_T= D_eT + \sigma^2_T (f=\infty)
\end{equation}
where $T$ is the integration time, $\sigma^2_T(f=\infty)$ is the count variance on the pixel 
corresponding to $T$, and $\sigma^2_T(f=\infty) \simeq R^2_e$. Dark current results are shown
in Table \ref{tab:darkcurrent}. As for the RON we considered the nominal
values $m = 600$ and $g = 1/30$, for all pixels. The estimated values are in  good
agreement with the ones provided by FLI.

\begin{table}[h]
\centering
\caption{Dark current for the three cameras.  First and second rows report the
values given by e2v (m = 1000) and First Light (m = 600) respectively.  The
values from our estimation (from binning 3x3 and 4x4 data) is shown in the  last
row. Notice that we considered an on-chip multiplication gain of 600 for  every
pixel.}\label{tab:darkcurrent}
\begin{tabular}{lccc}
	\toprule
	& & \bf{$D_e$ [e$^-$/pixel/s]} & \\
	\bf{Camera no.} &  
   	 \bf{39} &
   	 \bf{40} &
   	 \bf{41}\\
    \midrule
    \bf{e2v (m=1000, 1274fps)} & 5.10 & 6.37 & 5.10 \\
  	\bf{First Light (m=600, 2000fps)}	& 2.56 & 3.58 & 3.10  \\
  	\bf{OAA (m=600)} & 2.16 & 2.80 & 2.58  \\
    \bottomrule
  \end{tabular}
\end{table}

\section{Analysis of illuminated frames}

In this section we describe the measurements made with light and the following
data analysis. These measurements are used to compute the on-chip 
multiplication gain and electronic gain of the CCD. These measurements are  then
used to obtain a more accurate estimation of RON and dark current. Moreover, 
the spatial distribution of the gain will be used to evaluate the need of a 
flat field correction.

\subsection{Detector gain}\label{gain}
\begin{figure}[h!]
\centering
\subfigure[Variance vs counts for two pixels, $m=1$. Curves are the best fits (equation \ref{eq:sigma})\label{fig:var_count_m1}]
{\includegraphics[width=0.45\columnwidth]{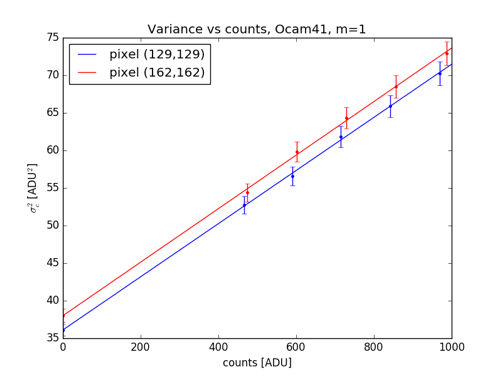}}
\subfigure[Variance vs counts for one pixel, $m=600$.\label{fig:var_count_m600}]
{\includegraphics[width=0.45\columnwidth]{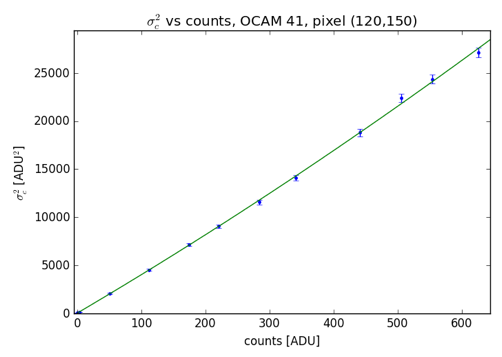}}
\subfigure[Total gain $G$ estimated as the linear component of equation \ref{eq:sigma}.
For each sector the three numbers are: the average value over the sector, the standard deviation
and the average fitting error.
\label{fig:tot_gain}]
{\includegraphics[width=0.45\columnwidth]{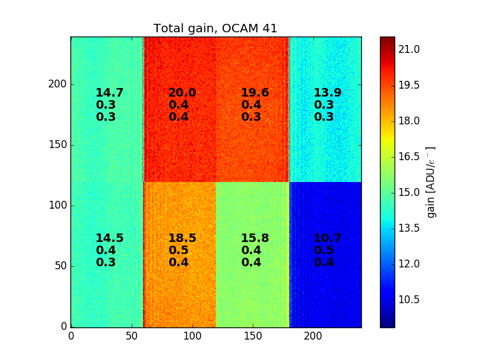}}
\subfigure[Quadratic term $k$ of equation \ref{eq:sigma}. 
For each sector the two numbers are: average value over the sector and standard deviation
of the quadratic part of the total gain.\label{fig:quad_term}]
{\includegraphics[width=0.45\columnwidth]{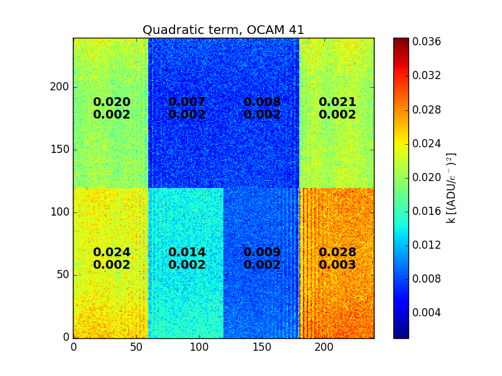}}
\caption{Typical output of the data analysis for camera no. 41.
}\label{fig:TypCamera41}
\end{figure}
\begin{table}[h]
\centering
\caption{Gains for camera no. 41 with $m = 600$ and all modes.
$k$ is the quadratic term and $G=m*g$ is the total gain
in ADU$/e^-$.}\label{tab:gains41}
\begin{tabular}{lcccccccccc}
	\toprule
	& \multicolumn{2}{c}{\bf{Bin 1}} & \multicolumn{2}{c}{\bf{Crop}} & \multicolumn{2}{c}{\bf{Bin 2}} &\multicolumn{2}{c}{\bf{Bin 3}} & \multicolumn{2}{c}{\bf{Bin 4}} \\
    \bf{Sector} & $k$ & $G[\frac{ADU}{e^-}]$ & $k$ & $G[\frac{ADU}{e^-}]$ & $k$ & $G[\frac{ADU}{e^-}]$ & $k$ & $G[\frac{ADU}{e^-}]$ & $k$ & $G[\frac{ADU}{e^-}]$ \\
    \midrule
    0 & 0.013 & 16.8 & -     & -    & 0.014 & 23.5 & 0.015 & 25.8 & 0.010 & 28.3\\ 
    1 & 0.004 & 21.5 & 0.005 & 21.4 & 0.004 & 28.3 & 0.001 & 31.6 & 0.000 & 32.9\\ 
    2 & 0.004 & 17.6 & 0.004 & 17.7 & 0.003 & 25.4 & 0.000 & 29.4 & 0.010 & 30.6\\ 
    3 & 0.014 & 13.4 & -     & -    & 0.016 & 18.4 & 0.016 & 22.0 & 0.002 & 24.8\\ 
    4 & 0.012 & 15.0 & -     & -    & 0.011 & 19.6 & 0.006 & 21.0 & 0.000 & 22.7\\ 
    5 & 0.003 & 19.7 & 0.004 & 19.7 & 0.002 & 24.9 & 0.000 & 25.2 & 0.000 & 25.5\\ 
    6 & 0.003 & 20.3 & 0.003 & 20.3 & 0.002 & 25.6 & 0.000 & 26.8 & 0.000 & 27.2\\ 
    7 & 0.014 & 14.8 & -     & -    & 0.014 & 19.3 & 0.012 & 20.7 & 0.008 & 22.6\\ 
    \bottomrule
  \end{tabular}
\end{table}

\begin{table}[h]
\centering
\caption{Gains for the three cameras in binning 1x1 mode.
$g$ is the electronic gain in ADU$/e^-$ (measured with $m=1$),
$k$ is the quadratic term and $G=m*g$ is the total gain
in ADU$/e^-$.}\label{tab:gainsAll}
\begin{tabular}{lccccccccc}
	\toprule
	& \multicolumn{3}{c}{no. 39, Bin1} & \multicolumn{3}{c}{no. 40, Bin1} & \multicolumn{3}{c}{no. 41, Bin1} \\
    \bf{Sector} & $g[\frac{ADU}{e^-}]$ & $k$ & $G[\frac{ADU}{e^-}]$ & $g[\frac{ADU}{e^-}]$ & $k$ & $G[\frac{ADU}{e^-}]$ & $g[\frac{ADU}{e^-}]$ & $k$ & $G[\frac{ADU}{e^-}]$ \\
    \midrule
    0 & 0.036 & 0.007 & 17.2 & 0.039 & 0.006 & 19.3 & 0.035 & 0.013 & 16.8 \\ 
    1 & 0.036 & 0.000 & 21.0 & 0.038 & 0.002 & 20.9 & 0.036 & 0.004 & 21.5 \\ 
    2 & 0.032 & 0.000 & 19.6 & 0.036 & 0.002 & 20.9 & 0.030 & 0.004 & 17.6 \\ 
    3 & 0.035 & 0.005 & 16.9 & 0.039 & 0.009 & 17.7 & 0.030 & 0.014 & 13.4 \\ 
    4 & 0.033 & 0.003 & 23.7 & 0.038 & 0.006 & 16.9 & 0.036 & 0.012 & 15.0 \\ 
    5 & 0.032 & 0.000 & 24.6 & 0.036 & 0.002 & 18.5 & 0.036 & 0.003 & 19.7 \\ 
    6 & 0.035 & 0.000 & 24.1 & 0.038 & 0.002 & 21.7 & 0.036 & 0.003 & 20.3 \\ 
    7 & 0.036 & 0.002 & 24.4 & 0.039 & 0.009 & 18.2 & 0.036 & 0.014 & 14.8 \\ 
    \bottomrule
  \end{tabular}
\end{table}
%
% \begin{table}[h]
% \centering
% \caption{Gains for camera no. 40 and 41 in Binning 2x2 mode.
% $g$ is the electronic gain in ADU$/e^-$ (measured with $m=1$),
% $k$ is the quadratic term and $G$ is the total gain in ADU$/e^-$.}\label{tab:gainsBin2}
% \begin{tabular}{lcccccc}
% 	\toprule
% 	& \multicolumn{3}{c}{no. 40, Bin2} & \multicolumn{3}{c}{no. 41, Bin2}  \\
%     \bf{Sector} & $g[\frac{ADU}{e^-}]$ & $k$ & $G[\frac{ADU}{e^-}]$ & $g[\frac{ADU}{e^-}]$ & $k$ & $G[\frac{ADU}{e^-}]$ \\
%     \midrule
%     0 & 0.058 & 0.009 & 23.2 & 0.057 & 0.014 & 23.5 \\ 
%     1 & 0.059 & 0.001 & 26.2 & 0.052 & 0.004 & 28.3 \\ 
%     2 & 0.056 & 0.000 & 26.6 & 0.052 & 0.003 & 25.4 \\ 
%     3 & 0.058 & 0.011 & 22.2 & 0.057 & 0.016 & 18.4 \\ 
%     4 & 0.055 & 0.006 & 20.4 & 0.051 & 0.011 & 19.6 \\ 
%     5 & 0.053 & 0.000 & 22.1 & 0.050 & 0.002 & 24.9 \\ 
%     6 & 0.054 & 0.000 & 27.6 & 0.049 & 0.002 & 25.6 \\ 
%     7 & 0.055 & 0.012 & 22.5 & 0.052 & 0.014 & 19.3 \\ 
%     \bottomrule
%   \end{tabular}
% \end{table}
%
We present here the measurements and analysis done to compute  on-chip
multiplication gain and electronic gain of the CCD.
At a given detector configuration, the fluctuation in time (frame by frame)  of
the pixel value is the sum of 
the Read Out Noise, $R_e$, and the photon noise, $\sigma_e^2$.
%several contributors; here we consider only three
%of them:
%\begin{equation}
%\sigma^2_T= R_e^2 + \sigma^2_e + K_e
%\end{equation}
%where $\sigma^2_T$ is the variance in time of a given pixel expressed in
%$e^{-^2}$, $R_e$ is the RON, $\sigma_e^2$ is the photon noise and $K_e$
%is an additional term to take into account possible non-linearities.
Since the random arrival rate of photons
controls the photon noise and the photon noise obeys the laws of Poissonian 
statistics we obtain:
\begin{equation}\label{eq:sigma}
\sigma^2_c= R_c^2 + \frac{G}{F}c + k c^2
\end{equation}
where $\sigma^2_c$ is the counts variance, $R_c=GR_e$ is Read Out Noise in square
counts, $c$ are the counts, $G=m*g$ is the total gain,
$k$ is an additional term to take into account possible non-linearities
and $F$ is the excess noise factor due to the electronic multiplication.
We considered $F=1$ for $m=1$ and $F=2$ for $m=600$.
Hence, without the electronic multiplication ($m= 1$, $F=1$),
and neglecting the non-linear term ($k<<10^{-3}$ for $m=1$), we obtain:
\begin{equation}\label{eq:var_counts}
\sigma^2_c= R_c^2 + gc
\end{equation}
We used an integrating sphere in order to acquire camera frames while the chip was 
illuminated uniformly. We used two different illumination fluxes and, for each 
one, we changed the integration time from the minimum up to 1ms. For each 
configuration (flux, $T$ and $g$) we acquire ~6000 frames.
We plot the counts variance versus the average counts  and we fit the data 
with equation \ref{eq:sigma} and \ref{eq:var_counts} for frames with m=600 and m=1 respectively.
The error on the variance ($\delta_{\sigma^2}$) is 
computed as:
\begin{equation}
\delta_{\sigma^2}= \sqrt{\frac{1}{n}\left(\mu_4 - \frac{n-3}{n-1}\mu_2^2\right)}
\end{equation}
\begin{equation}
\mu_k = \frac{1}{n-1}\sum_{i=1}^{n}(x_i-\bar{x})^k
\end{equation}
where $\mu_k$ is the $k$-th central moment, $x_i$ is the $i$-th value and
$\bar{x}$ is the mean on the events.
The additional term $kc^2$ introduces nonlinear behavior when $m>1$. 
The nonlinear term follow approximatively a quadratic trend respect to the 
average counts, as will be detailed in this section.\\
Figure \ref{fig:TypCamera41} reports the typical output of the data analysis for camera no. 41.
The figures refer to data in mode 1 (binning 1x1), with multiplication $m$ = 1 
(Figure \ref{fig:var_count_m1}) and $m$ = 600 (Figure \ref{fig:var_count_m600},
\ref{fig:tot_gain} and \ref{fig:quad_term}).\\
According with the specification from First Light the electronic gain of the 
CCD should be $g=0.03$ADU$/e^{-}$ and with $m$ = 600, we have $G = m * g =$ 20 ADU/$e^-$ .
The values measured in the central part of the CCD (quadrants 1, 2, 5, 6) are in 
good agreement with the expected value and the quadratic term is negligible 
(Figure \ref{fig:TypCamera41}). In the external part of the CCD (quadrants 0, 3, 4, 7) the 
quadratic term is more relevant and the total gain is depressed. Camera no. 41 
measurements show bigger quadratic terms respect to the other two cameras (see
Table \ref{tab:gainsAll}).\\
Table \ref{tab:gains41} reports the results of the analysis on camera no. 41
with multiplication set to 600.\\
We can notice that the total gain becomes bigger rising the binning
(see Section \ref{sec:Ron2} to further considerations),
and highest values of the total gain are associated to the lowest quadratic terms
(see the results for camera n.41 in Table \ref{tab:gainsAll}).% and \ref{tab:gainsBin2}).

\subsection{Read-Out Noise and Dark current with measured gains}\label{sec:Ron2}

Read-Out Noise and Dark current values shown in the previous section are computed using nominal
values for $g$ and $G$.
We recomputed RON and dark estimates (section \ref{sec:Dark}), using the measured values of $g$ and $G$ found in section \ref{gain}.
In Table \ref{tab:gains41} we see that the $G$ values has a
positive trend with the binning, reaching at bin 4 a $+25 \div 50\%$ of its value at bin 1.
Considering this effect, the RON results reduced for binned modes ($<0.5e^-$ at all binning) and the estimation of $D_e$ is consistent at all binning. As example, in Table \ref{tab:darkcurrentUpdated} we report for camera n.41 the comparison for $R_e$ and $D_e$ between values obtained with nominal and measured $g$ and $G$.
%gains of the camera, but if we update these values with the measured gains
%we get smaller RON values for higher binning and almost uniform dark current values
%(see Table \ref{tab:darkcurrentUpdated}),
%because $G$ slightly increases with binning (see Table \ref{tab:gains41}).
\begin{table}[h]
\centering
\caption{RON and dark current for camera no. 41 computed using measured gains
from Table \ref{tab:gains41}.
Dark current values always refer to physical pixel (not binned one).}
\label{tab:darkcurrentUpdated}
\begin{tabular}{lcccc}
	\toprule
	& \multicolumn{2}{c}{\bf{$R_e$ [e$^-$]}} & \multicolumn{2}{c}{\bf{$D_e$ [e$^-$/pixel/s]}}\\
	\bf{Camera no.} &  \multicolumn{4}{c}{\bf{41}}\\
    \bf{gain value} & \bf{nominal} & \bf{measured} & \bf{nominal} & \bf{measured}\\
    \midrule
  	\bf{bin1x1} & 0.34 & 0.40  & 1.16 & 1.46\\
  	\bf{bin2x2} & 0.47 & 0.38  & 2.18 & 1.60\\
  	\bf{bin3x3} & 0.56 & 0.45  & 2.43 & 1.47\\
  	\bf{bin4x4} & 0.67 & 0.51  & 2.58 & 1.38\\
    \bottomrule
  \end{tabular}
\end{table}

\section{Gain uniformity}\label{flatField}
%
%\begin{figure}[h!]
%\centering
%\includegraphics[width=0.55\columnwidth]{immagini/fps.png}
%\caption{
%Average counts on the columns with uniform illumination (about 100counts) over
%half the frame. The second half presents similar behavior. The values are 
%normalized by the average on the sector.  Different colors represent different 
%framerates.}
% \label{fig:fps}
%\end{figure}
%
%
Detector gain is a key aspect for PWFS because, in order to make a correct measurements,
it requires a spatially uniform response on the chip regions where the
pupil images lie.
%good flat field calibration on the entire chip.
In fact, PWFS slopes calculation is done comparing the amount of light in the
four pupil images\cite{pyramid}: 
different gains in the regions where the pupils light fall
yields to an error on the flux measurement,
propagating it in the slope estimation. 

\begin{figure}[h!]
\centering
\subfigure[Average counts on the rows with uniform illumination ($\sim$100counts) over half the 120$\times$120 central sub-frame. The values are normalized by the average on the sector.  Different colors represent different frame rates.\label{fig:fps}]
{\includegraphics[width=0.45\columnwidth]{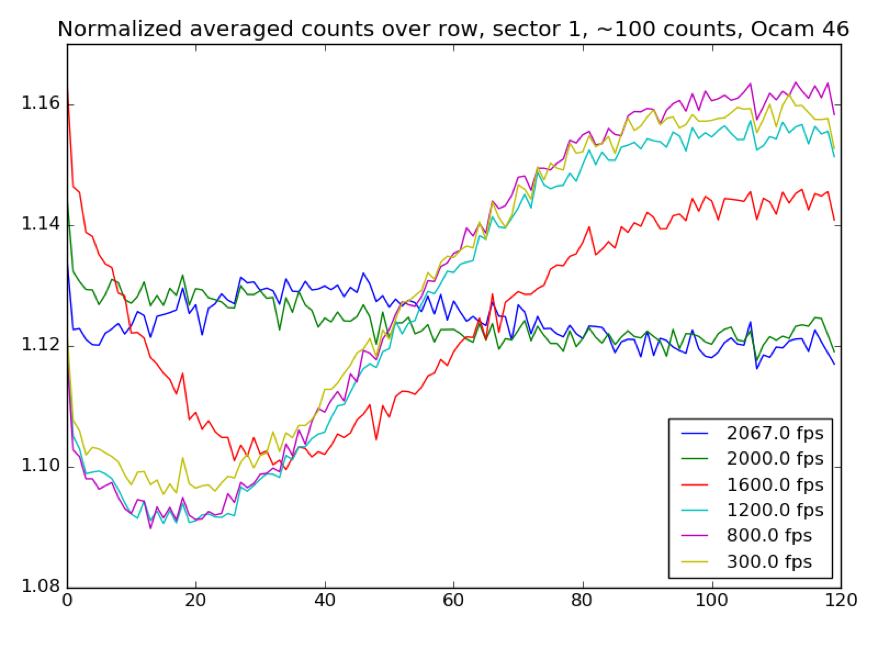}}
\subfigure[Average counts over the rows for the sectors in the bottom part of the chip (from 0 to 3). Blue line is wit no flat fielding, green line flat fielding with a previous power cycle and red line is flat fielding from a previous power cycle with the sector by sector re-normalization.\label{fig:FlatFieldCorrection}]
{\includegraphics[width=0.45\columnwidth]{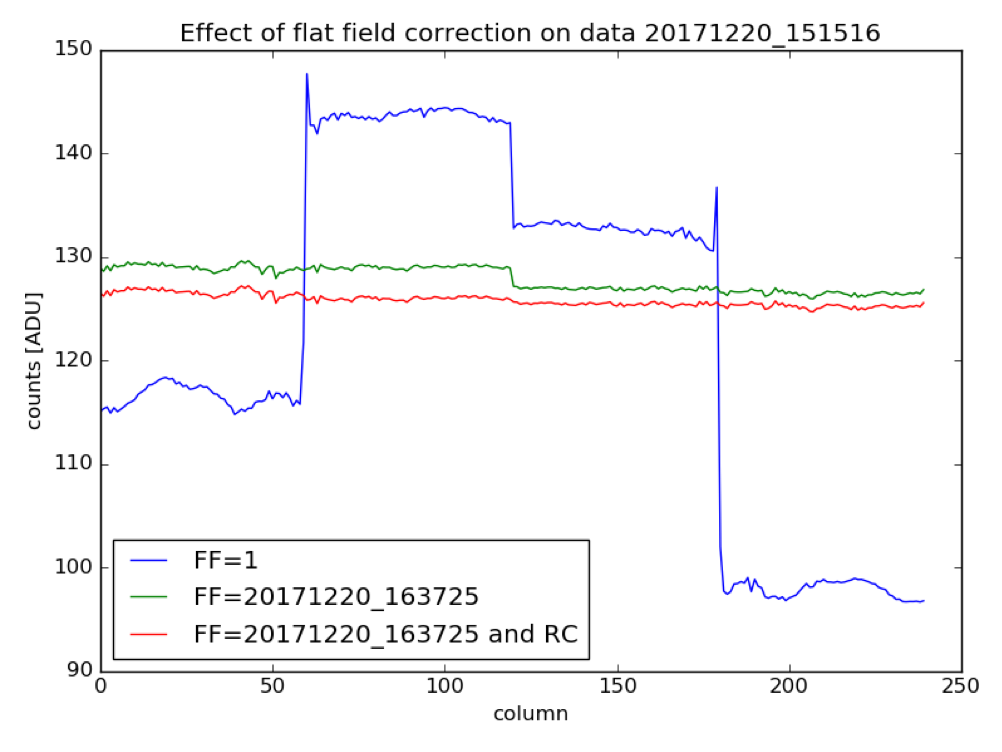}}
\caption{Flat field analysis.}
 \label{fig:flatFieldAnalysis}
\end{figure}

As can be seen in Figure \ref{fig:tot_gain} (and Figure \ref{fig:FlatFieldCorrection} blue line, Table \ref{tab:gains41} and \ref{tab:gainsAll}), Ocam2k gain $G$, when EM multiplication is enabled, varies manly sector by sector in the range $\pm25\%$.
Hence, flat fielding is required to work properly with PWFS
\footnote{Notice that FLAO system\cite{FLAO} does not use flat field correction because
e2v CCD39 has a spatially uniform gain.}
and, in this section, we present our analysis on the stability
of the camera gains needed to set up a correct
procedure to calibrate and manage this flat fielding.
Flat field measurement has been performed using an integrating sphere in order to provide uniform illumination on the chip.
In the following part of this section we will only refer to the four central sectors of the chip,
because the PWFS of the SOUL project has pupils which lie on 
a region of interest of 120$\times$120pixels.

We will focus on two main dependencies of the flat field found during laboratory analysis: camera power cycle and integration time.
After a power cycle of the camera, the average $G$ value typically changes
sector by sector of few percents (see green line of Figure \ref{fig:FlatFieldCorrection}).
This effect can be compensated by measuring the mean $G$ value for each sector,
at each power cycle, and then re-normalizing the flat field sector by sector with the obtained values.
The mean $G$ value can be measured
averaging the ratio between mean counts and
time variance when some illumination is present.
Using this procedure we need a light source stable during data acquisition (few seconds)
without any requirements of spatial uniformity.
Applying this method, the variation between sectors goes below 1\% (see red line of Figure \ref{fig:FlatFieldCorrection}).
The second feature we are reporting here is the gain change row by row as function of the 
frame rate. In Figure \ref{fig:fps} we report the counts averaged over the rows (with spatially uniform illumination) at different frame rates when both the flat field (acquired at maximum frame rate) and the sector by sector correction are applied. In the frame rate range 300-2000Hz, the maximum variation of the gain across the frame has a peak-to-valley $<10\%$.
A similar behavior is already described in literature
(section 4 of Dowing \emph{et al.}\cite{Dowing2013}).
This effect can be mitigated measuring and applying different flat fields for different frame rates reducing the gain variation to a few $\%$.

\subsection{Impact of gain non uniformity on SOUL performances}

This section is aimed to quantify the impact of the camera gain features
described in the previous section, hypothesizing to not update flat field
in case of framerate variations and camera power cycles.
This estimation has been carried on using 
PASSATA\cite{doi:10.1117/12.2233963}, the end-to-end simulator used for all numerical
simulations in the SOUL project. We run full adaptive optics end-to-end simulations of the SOUL system, using different detector gain maps, emulating the effects described in sect. \ref{flatField}. Simulating the SOUL system, we considered an OCam2k sub-frame of 120 x  120 pixels over the 4 central sectors. The AO system calibration (interaction matrix corrector-WFS) is simulated applying a flat
detector gain map. Then the AO simulations are run using the following gain maps:

\begin{enumerate}[(A)]
\item reference case: perfect flat field, all normalized gains are 1; 
\item power cycle effect: 3 sectors gain value 1, fourth sector with gain 1.1;
\item frame rate effect: we used real averaged frames acquired at 1200Hz normalized with a flat field acquired at 2000Hz.
\end{enumerate}

\begin{table}[h]
\centering
\caption{Simulation parameters.}
\label{tab:parameters}
\begin{tabular}{llllll}
	\toprule
	\bf{atmosphere} & \bf{seeing}     & 0.8arcsec\\
    		        & \bf{$L_0$}      & 40m\\
                    & \bf{aver. wind speed} & 16m/s\\
    \midrule
    \bf{$1^{st}$ configuration} & \bf{R magnitude}    & 8 &
    \bf{$2^{nd}$ configuration} & \bf{R magnitude}    & 13.5\\
    	      & \bf{mod. amplitude} & $\pm 3\lambda/D$ &
   	          & \bf{mod. amplitude} & $\pm 4\lambda/D$ \\
              & \bf{sampl. freq.}   & 1500Hz &
              & \bf{sampl. freq.}   & 300Hz\\
              & \bf{corr. modes}    & 630 &
              & \bf{corr. modes}    & 299\\
              & \bf{total time}     & 10s &
              & \bf{total time}     & 16s\\
    \bottomrule
  \end{tabular}
\end{table}

We simulated standard seeing conditions at LBT ($0.8$arcsec) and two cases of natural guide star: bright end (R=8, 630 corrected modes) and medium-faint (R=13.5, 299 corrected modes); more simulation parameters are reported in Table \ref{tab:parameters}. In Table \ref{tab:results} we summarize the simulation results in terms of total wavefront residual for the considered cases. In the two right columns of the table we can see the residual wavefront difference for cases B and C with respect to the reference case, including or not the tip-tilt. These numbers show, as expected, that case B introduces mainly a static tilt, with negligible effects on higher modes. The amount of introduced tilt depends on the reference star flux. About case C, it affects a wide range of modes but with negligible impact on the overall system performances.

\begin{table}[h]
\centering
\caption{Simulation results.}
\label{tab:results}
\begin{tabular}{lccccc}
	\toprule
	\bf{Case} & \bf{conf.} &  \bf{res. RMS [nm]} & \bf{diff. [nm]} & \bf{diff. w/o TT [nm]}\\
    \midrule
    A &  $1^{st}$ &  87.1 & - & -\\
    B &  $1^{st}$ & 95.9 & 40.1 & 4.0\\
    C &  $1^{st}$ &  87.2 & 3.8 & 3.8\\
    A &  $2^{nd}$ &  179.0 & - & -\\
    B &  $2^{nd}$ &  192.2 & 70.0 & 8.8\\
    C &  $2^{nd}$ &  182.5 & 35.4 & 31.8\\
    \bottomrule
  \end{tabular}
\end{table}

Hence, we can conclude that:
\begin{itemize}
\item the power cycle effect (case B) introduces a static tilt of the order of $100$nm RMS, translating into an indetermination of the scientific PSF position of about 10mas\footnote{here we consider a pupil with 8.2m diameter as for the SOUL at the LBT.}; compensating with sector gain corrections, as mentioned in \ref{flatField}, we expect to reduce this effect down to the order of a few mas;
\item the frame rate effect has already a negligible impact in terms of Strehl ratio in NIR bands; using a set 3-4 flat fields, calibrated at different camera frame rates, this effect can be easily further reduced for high contrast applications.  
\end{itemize}

\section{Conclusion}

Our laboratory measurements show that the 3 considered Ocam2k cameras are compliant with their specifications in terms of RON (about 0.4e- and 0.4-0.7e- for binned modes) and dark current (1.5e-/pixel/s). We estimated, pixel by pixel, the camera gain ($g$) and the multiplication gain ($m$) obtaining values close to the nominal ones but showing differences sector by sector. The 4 internal sectors of the chip show better values for $g$ and $G= g*m$, and a more linear behavior of the relation counts vs. time variance of counts. We found that the camera gain varies with power cycles and frame rate. Via numerical simulations we quantified the impact of these variations on the adaptive optics performances. The power cycle changes the gain sector by sector and, in the SOUL case, this translates into a global static tilt of about 10mas that can even be mitigated with an initial measurement of the sector gains. On the other side, the gain variation associated to frame rate change has a minimal impact on the performances.

The measurements and the simulations we performed do not show any major issue for the use of Ocam2k for high order PWFSs when pupil images are allocated one per chip sector, as in the SOUL case. In the case of ELTs (PWFS for GMT\cite{doi:10.1117/12.2057059} and EELT\cite{2018arXiv180700073D,2016SPIE.9909E..0AC}), the pupil images will not fit into a single 60x120 pixel sector and the impact of the gain variations will be higher. Of course, each ELT case deserves dedicated numerical simulations; however, from the described experience with SOUL simulation and cameras, we are inclined to think that the gain effect can be efficiently mitigated with dedicated calibrations, allowing the use of Ocam2k even for the ELT PWFSs.

\bibliography{biblio}
\bibliographystyle{spiebib}

\end{document}